\documentstyle[12pt]{article} 
\title{Non-Universal Critical Behaviour \\ of 
             \\ Two-Dimensional Ising Systems} 
\author{{\sc Kazuhiko Minami and Masuo Suzuki*} \\ \\
  {\small{\it Magnetic Materials Laboratory,}} \\
  {\small{\it The Institute of Physical 
              and Chemical Research (RIKEN),}}\\
  {\small{\it  Wako, Saitama, 351-01, JAPAN}}\\
  {\small{\it *Department of Physics, 
               University of Tokyo,}}\\
  {\small{\it  Hongo, Bunkyoku, Tokyo, 113, JAPAN}}} 
\date{PACS number: 05.50}

\newcommand{\beq}{\begin{equation}}
\newcommand{\eeq}{\end{equation}}
\newcommand{\beqa}{\begin{eqnarray}}
\newcommand{\eeqa}{\end{eqnarray}}
\newcommand{\bit}{\begin{itemize}}
\newcommand{\eit}{\end{itemize}}
\newcommand{\bdes}{\begin{description}}
\newcommand{\edes}{\end{description}}
\begin{document} 
\maketitle 
\begin{center}
Abstract
\end{center} 

Two conditions are derived for Ising models 
to show non-universal  critical behaviour, 
namely conditions concerning 
1) logarithmic singularity of the specific heat and 
2) degeneracy of the ground state. 
These conditions are satisfied with the eight-vertex model, 
the Ashkin-Teller model, 
some Ising models with short- or long-range interactions 
and even Ising systems without 
the translational or the rotational invariance. 

\newpage 
\section{Introduction}
\setcounter{equation}{0}
The universality of critical exponents is 
one of the most important concepts 
in critical phenomena. 
According to this universality hypothesis, 
critical exponents depend only upon 
the dimensionality, the symmetry 
and the interaction range of Hamiltonians, 
namely they are independent on 
the details of the relevant Hamiltonian 
such as the strength of the interactions 
in ordinary situations. 

It is quite interesting to study 
in what condition this hypothesis is violated 
from the form of the Hamiltonian. 

The first example violating the universality hypothesis is 
the eight-vertex model solved by Baxter\cite{baxter}. 
This model can be mapped\cite{wu}\cite{kadweg} 
onto the two-layered square-lattice Ising model 
in which two layers interact each other 
via a four-body interaction $J_4$. 
The critical exponents of this model vary with 
the parameter $\mu$ 
which is a function of interaction energies. 
The exponents are obtained\cite{baxter} as 
$\alpha=2-\pi/\mu$, $\beta=\pi/16\mu$, $\nu=\pi/2\mu$, 
and the scaling hypothesis 
or the weak universality hypothesis\cite{weak} 
insists that 
$\gamma=7\nu/4=7\pi/8\mu$, $\delta=15$, $\eta=1/4$. 
Kadanoff and Wegner\cite{kadweg} 
have shown that the existence of 
marginal operators is a necessary condition 
for appearance of 
continuously varying critical exponents. 
Kadanoff and Brown\cite{kadbrw} 
have shown that the long-range behaviour of 
the correlations of the eight-vertex 
and the Ashkin-Teller models are 
asymptotically the same as those of the Gaussian model. 

There exists another model 
which consists of short-range two-body 
interactions and which is believed 
to have continuously varying critical exponents. 
The $s=1/2$ square-lattice Ising model 
with the nearest-neighbour 
interaction $J$ and the next-nearest 
neighbour interaction $J'$ have been studied by many 
authors\cite{saf1}-\cite{saf*} to obtain 
the phase diagram. 
The numerical calculations of van Leeuwen,\cite{leeu} 
Nightingale\cite{nigh} 
and Swendsen and Krinsky\cite{swkr} 
are the first to show non-universal critical behaviour 
of this model. 
The singular part of the free energy 
is calculated perturbatively 
by Barber\cite{barber}. 
The results of the high temperature expansion 
by Oitmaa\cite{oit} 
agrees with these calculations. 
The coherent-anomaly method (CAM) 
is applied\cite{cam1}-\cite{saf4b} 
to this model and the continuously 
varying critical exponents are estimated 
with errors smaller than $\sim 1\%$\cite{saf}\cite{saf4b}. 
The degeneracy of the ground state energy 
and the existence of a multi-component order parameter 
have been studied by J\"{u}ngling\cite{jung} 
and by Krinsky and Muhamel\cite{krin}. 
A more general Hamiltonian has been investigated 
by the present authors\cite{saf4b} 
which includes the above two models as special cases of it  
and which has continuously varying critical exponents. 

In the present paper, 
we phenomenologically derive a sufficient condition 
to have continuously varying critical exponents. 
This study is the generalization of the argument 
reported in ref.\cite{lett}. 
Our condition is satisfied in the eight-vertex model, 
the Ashkin-Teller model 
and the $s=1/2$ square-lattice 
Ising model with the next-nearest-neighbour interaction. 
Our condition is also satisfied with some systems 
including long-range interactions 
and with some systems 
without the translational or the rotational invariance. 

Brief explanations of the relevant models 
are given in section 2, 
together with some explicit conditions 
on continuously varying critical exponents. 
A phenomenological perturbation scheme\cite{super} 
is explained in section 3 
and applied to the eight-vertex model 
in order to demonstrate that 
our scheme provides the exact first-order derivative 
of the critical temperature 
and the exponent $\gamma$ 
of the eight-vertex model in section 4. 
The temperature dependence of correlation functions  
and the characteristic cancellations 
of interaction energies at the ground state 
are discussed in section 5. 
The symmetry of the relevant model is studied in section 6. 
The perturbational scheme explained in section 3 is applied, 
in section 7, 
to the model defined in section 2. 
It is derived that 
there exist finite first- or second-order derivatives 
(with respect to the interaction energies) 
of the critical exponent $\gamma$ 
and hence it varies continuouly 
as a function of interaction energies. 
Finally, the condition in section 2 is generalized 
to include a more general type 
of interactions in section 8. 

\section{Models}
\setcounter{equation}{0}
Let us consider the following Hamiltonian 
\beq
{\cal H}=\sum_k{\cal H}_k+\sum_{kl}{\cal H}_{kl} 
\label{genham}
\eeq
where $\{{\cal H}_k\}$ 
is a finite set of two-dimensional Ising systems. 
In the present paper, we derive that the model 
with the Hamiltonian (\ref{genham}) 
have continuously varying critical exponents  
provided it satisfies 
the following two conditions 1) and 2). 

\bdes 
\item[1)] 
The specific heat of ${\cal H}_k$ 
shows the logarithmic singularity 
at the critical temperature $T_{\rm c}$ 
which is independent on $k$. 
\item[2)]
The ground state energy 
of the total Hamiltonian ${\cal H}$ 
is invariant for the spin inversion of each ${\cal H}_k$.  
Here ${\cal H}_{kl}$ is written as  
\beq
{\cal H}_{kl}
=-J_{kl}\sum_i{\cal O}^{(k)}_{i}{\cal O}^{(l)}_{i}, 
\eeq
where ${\cal O}^{(k)}_{i}$ and ${\cal O}^{(l)}_{i}$
 are the $n^{(k)}$-body and 
$n^{(l)}$-body spin product of spins 
belonging to ${\cal H}_k$ and ${\cal H}_l$, 
respectively. 
\edes

The logarithmic singularity in the condition 1) 
results in the temperature 
dependence of the correlation functions of the form  
$\langle s_is_j\rangle=c_0+c_1\epsilon\ln\epsilon$, 
where $\epsilon=(T-T_{\rm c})/T_{\rm c}$ and $c_0$ and $c_1$ 
are some constants. 
This formula is derived in section 5. 

The restriction on ${\cal O}^{(k)}_{i}$ 
and ${\cal O}^{(l)}_{i}$ 
in the condition 2) can be partially removed 
and more complicated interactions are permitted, 
i.e. $n^{(k)}$ and $n^{(l)}$ 
can depend on the region ${\cal R}_m$ 
in (\ref{grncan}). 
This generalization is explained in section 8. 

\section{Perturbational scheme}
\setcounter{equation}{0}
To derive non-universal critical behaviour of these models, 
let us consider the perturbational expansion\cite{super} 
with respect to interaction energies. 
The susceptibility $\chi$ is assumed to behave as 
\beq
\chi\sim\epsilon(J)^{-\gamma (J)}, 
\;\;\epsilon(J)=(T-T_{\rm c}(J))/T_{\rm c}(J), 
\label{dchieps}
\eeq
where $J$ is an interaction energy. 
Differentiating it with respect to $J$, we obtain 
\beq
(\frac{\partial \chi}{\partial J})_{J=0}\simeq
\chi_0 
[-\gamma (0)\frac{1}{\epsilon(0)}
  (\frac{\partial\epsilon}{\partial J})_0 
 -(\frac{\partial\gamma}{\partial J})_0 \log\epsilon(0)]
\label{dchidif}, 
\eeq
where the subscript $0$ denotes $J=0$. 
Hence we can obtain $(\partial\gamma/\partial J)_0$ 
from the coefficient of 
the temperature dependence $\chi_0 \log\epsilon$. 
On the other hand, 
the susceptibility is expressed by the two-spin 
correlation functions in the form 
\beq
\chi=\beta\mu_{\rm B}^2\sum_{i_0j_0} g_{i_0j_0}
\langle s_{i_0}s_{j_0}\rangle, 
\label{chicor}
\eeq
where $\beta=1/k_{\rm B} T$, $\mu_{\rm B}$ 
is the Bohr magneton  
and $\{g_{i_0j_0}\}$ 
denote the signs coming from the emerging order. 
We differentiate (\ref{chicor}) and estimate 
$(\partial\gamma/\partial J)_0$ 
by comparing with (\ref{dchidif}). 
The existence of a finite and non-vanishing derivative 
of the exponent $\gamma$ 
is an evidence 
for continuously varying critical exponents 
to appear. 

\section{Example}
\setcounter{equation}{0}
Here we give two important examples. 
Let us consider the zero-field eight-vertex model. 
The Hamiltonian ${\cal H}_{8V}$ 
of this model is written as 
${\cal H}_{8V}={\cal H}_1+{\cal H}_2+{\cal H}_{12}$, 
where ${\cal H}_1$ and ${\cal H}_2$ 
denote the following Hamiltonians 
\beqa
{\cal H}_1&=&-J'\sum_{i+j=
{\rm even}}(s_{ij}s_{i+1j+1}+ s_{ij}s_{i+1j-1}),
\nonumber\\
{\cal H}_2&=&-J'\sum_{i+j=
{\rm odd }}(s_{ij}s_{i+1j+1}+ s_{ij}s_{i+1j-1}),  
\label{ham12}
\eeqa
and ${\cal H}_{12}$ is  
\beqa
{\cal H}_{12}&=&-J_4\sum s_{ij}s_{i+1j+1}s_{i+1j}s_{ij+1}. 
\eeqa
This model decouples into 
the two square-lattice Ising models 
${\cal H}_1$ and ${\cal H}_2$  when $J_4=0$. 
The interaction ${\cal H}_{12}$ has even symmetry 
for the spin inversion 
of each subsystem ${\cal H}_1$ and ${\cal H}_2$ 
and hence the energy of ${\cal H}_{8V}$ 
is four-fold degenerated 
in the whole temperature region. 
This model obviously satisfies 
the conditions 1) and 2) in section 2. 
The weight $g_{i_0j_0}$ equal $1$ for $J'>0$. 
Differentiating (\ref{chicor}) with respect to $J_4$ 
and using the fact 
that the model decouples into two layers when $J_4=0$, 
we obtain the following expression 
\beq
(\frac{\partial\chi}{\partial J_4})_{J_4=0}
=\beta^2\mu_{\rm B}^2\sum_{i_0j_0ijkl}[
 \langle s_{i_0j_0}s_{ij}s_{kl}s_{k+1l+1}\rangle_0 
-\langle s_{i_0j_0}s_{ij}\rangle_0
 \langle s_{kl}s_{k+1l+1}\rangle_0 ]
\langle s_{kl+1}s_{k+1l}\rangle_0, \label{dchi8V}
\eeq
where $\langle\;\rangle_0$ 
denotes the expectation value for $J_4=0$.  
This expression coincides with 
\beq
(\frac{\partial\chi}{\partial J'})_{J_4=0}\times\omega_0
=\frac{\chi_0}{\epsilon(0)}\frac{\gamma(0)}{J'}\omega_0. 
\label{ex}
\eeq
Here $\omega_0$ denotes 
the nearest-neighbour two-spin correlation function 
which shows the following behaviour 
\beq
\omega_0\simeq\frac{1}{\sqrt{2}}
       +\frac{4J'}{\pi k_B T_c(0)}\epsilon(0)\log\epsilon(0) 
\label{twocor}
\eeq
at the critical point. 
Comparing (\ref{ex}) and (\ref{twocor}) with (\ref{dchidif}), 
we obtain 
\beq
(\frac{\partial T_c  }{\partial J_4})_{J_4=0}=
\frac{T_c(0)}{\sqrt{2} J'} 
\ \ {\rm and}\ \ 
(\frac{\partial\gamma}{\partial J_4})_{J_4=0}=
                      -\frac{4\gamma(0)}{\pi k_B T_c(0)},
\label{dgam}
\eeq
which are identical with the derivatives 
obtained from the exact result by Baxter. 

The square-lattice Ising model 
with antiferromagnetic next-nearest-neighbour 
interactions is another example. 
The Hamiltonian of this model is given in the form 
${\cal H}_{\rm SAF}={\cal H}_1+{\cal H}_2+{\cal H}_{12}$, 
where ${\cal H}_1$ and ${\cal H}_2$ 
are given by (\ref{ham12}) for $J'<0$   
and ${\cal H}_{12}$ is  
\beqa
{\cal H}_{12}&=&-J\sum_{ij}( s_{ij}s_{i+1j}+s_{ij}s_{ij+1}). 
\eeqa 
The ground state of this model 
is ordered as the N\'{e}el state in each sublattice 
for the interaction region $|J/J'|<2$, 
in which the ground state energy is 
invariant for the spin inversion of each sublattice. 
It is in this interaction region $|J/J'|<2$ 
that this model is considered to have 
continuously varying critical exponents. 
This model also satisfies the condition in section 2. 
It is derived in ref.\cite{barber} and \cite{lett} 
that this model has continuously varying critical exponents 
with the derivatives $(\partial\gamma/\partial J)_0=0$ 
and $(\partial^2\gamma/\partial J^2)_0\neq 0$. 

\section{Preliminary formulas}
\setcounter{equation}{0}
The temperature dependence of correlation functions 
is specified from the 
condition 1).
For the purpose to obtain multi-spin correlation functions, 
let us consider 
${\tilde{\cal H}}_k={\cal H}_k-{\tilde J}{\cal O}$, 
where ${\cal O}$ denotes a certain spin product. 
The free energy ${\tilde f}_{ks}$ of this Hamiltonian 
is differentiable with respect to ${\tilde J}$. 
Assuming that ${\tilde f}_{ks}$ 
shows logarithmic or power-law behaviour for
${\tilde J}=0$,  it should have the form 
\beq
{\tilde f}_{ks}=C_1\epsilon^{2-\alpha_1}\log\epsilon+C_2\epsilon^2
\frac{1-\epsilon^{-\alpha_2}}{\alpha_2},\label{3free}
\eeq
where $C_1$ and $C_2$ are some constants, 
$\alpha_1$ and $\alpha_2$ are functions of 
${\tilde J}$ with $\alpha_1({\tilde J}=0)=0$ 
and $\alpha_2({\tilde J})\rightarrow 0$ 
for ${\tilde J}\rightarrow 0$. 
The second term converges to $C_2\epsilon^2\log\epsilon$ 
when ${\tilde J}\rightarrow 0$. 
From (\ref{3free}), 
\beq
\langle {\cal O}\rangle
=\frac{1}{\beta}
(\frac{\partial{\tilde f_{ks}}}{\partial{\tilde J}})
_{{\tilde J}=0}
\propto 2\epsilon
(\frac{\partial\epsilon}{\partial{\tilde J}})_0\log\epsilon
+O(\epsilon), 
\eeq
where contributions vanishing faster than $\epsilon\log\epsilon$ 
(for $\epsilon\rightarrow 0$) 
are included in $O(\epsilon)$. 
The regular part
of the free energy yields  a certain constant term. 
Then we generally obtain 
the temperature dependence of correlation functions 
in the form  
\beq
\langle{\cal O}\rangle\cong c_0+c_1\epsilon\log\epsilon. 
\label{gencor}
\eeq
 
The condition 2) results in the strong cancellations 
of interaction energies at the ground state. 
Here we introduce notations for the ground-state configuration 
of the system as $\{g_{i}\}$ 
and the spin product ${\cal O}_i=s_{i_1}\cdots s_{i_n}$ 
in ${\cal H}_{kl}$ at the ground state as 
${\cal G}_{i}=g_{i_1}\cdots g_{i_n}$. 
The $J_{kl}$-dependent part of the ground state energy, 
$\epsilon_G(J_{kl})$, is written as 
\beq
\epsilon_G(J_{kl})=-J_{kl}\sum_{i}{\cal G}_{i}.
\eeq
In the case where $\sum_{i}{\cal O}_{i}$ 
has even symmetry for the spin 
invergion of  ${\cal H}_k$ and ${\cal H}_l$ 
(i.e. both $n^{(k)}$ and $n^{(l)}$ are even), 
the condition 2) is automatically satisfied. 
Otherwise, the condition 2) yields 
\beq
-J_{kl}\sum_{i}{\cal G}_{i}=
-J_{kl}\sum_{i}(-{\cal G}_{i}) 
\ {\rm that is}\  
\sum_{i}{\cal G}_{i}=0. \label{cangrn}
\eeq

We exclude, from our arguments, 
the case that the condition 2) is asymptotically 
satisfied only in the thermodynamic limit. 
Then we can rewrite the condition (\ref{cangrn}) as 
\beq
\sum_{i}{\cal G}_{i}=
\sum_m\sum_{i\in{\cal R}_m}{\cal G}_{i},\ \ {\rm and}\ \ 
\sum_{i\in{\cal R}_m}{\cal G}_{i}=0, \label{grncan}
\eeq
where $\{{\cal R}_m\}$ 
denote a set of finite regions containing 
a finite number of spins  
and the symbol $i\in{\cal R}_m$ 
denotes that all the spins in ${\cal G}_i$ are included 
in ${\cal R}_m$. 

\section{Symmetries and the vanishing derivatives}
\subsection{Basic symmetries}
\setcounter{equation}{0}
In this section, we derive some properties 
which depend only on the symmetry 
of the relevant model. 
We consider here the case $N=2\ $ 
i.e., the following Hamiltonian 
${\cal H}={\cal H}_1+{\cal H}_2+{\cal H}_{12}$ 
for $\epsilon >0$. 

We assume that ${\cal H}_{12}$ has odd symmetry 
for the spin invergion of ${\cal H}_2$. 
Then the interaction ${\cal H}_{12}$ has the form 
\beq
{\cal H}_{12}=-J\sum {\cal O}^{(1)}_{i}{\cal O}^{(2)}_{i}, 
\eeq
where ${\cal O}^{(2)}_{i}$ 
changes its sign for the spin inversion 
of ${\cal H}_2$. 
Hereafter we omit the subscript of $J_{12}$ 
and write $J_{12}$ as $J$ for simplicity. 

Let us consider the transformation 
of spin configurations 
in which all the spins on ${\cal H}_2$ are reversed 
and the spins on ${\cal H}_1$  are unchanged. 
The energy of the model ${\cal H}$ 
with the interaction $J$ for the spin 
configuration $C$ equals to the energy 
of the model ${\cal H}$ 
with the interaction $-J$ 
for the transformed spin configuration $C'$. 
As a result, 
we can find  one-to-one correspondence 
of the Boltzmann factor 
coming from the configuration $C$ and $C'$. 
Summing over all configurations, 
we find that the partition functions $Z(J)$ 
and $Z(-J)$ are the same 
\beq
Z(J)=Z(-J). \label{Z} 
\eeq
Since we consider the partition function 
without an external field, 
we obtain from (\ref{Z}) 
for the critical temperature $T_{\rm c}(J)$ and 
the exponent $\alpha(J)$ of the specific heat as 
\beq
T_{\rm c}(J)=T_{\rm c}(-J)\ \ {\rm and}\ \ 
\alpha(J)=\alpha(-J). \label{alpha} 
\eeq
From (\ref{alpha}) we can conclude that 
any odd derivatives 
of $T_{\rm c}(J)$ and $\alpha (J)$ 
are equal to zero when they exist. 

Next we study the symmetry of the correlation functions 
$\omega_{i_0j_0}(J)$ 
which are the expectation values 
of the following two-spin product with the weight 
corresponding the emerging order as  
\beq
\omega_{i_0j_0}(J)\equiv
\frac{{\rm Tr}g_{i_0j_0}s_{i_0}s_{j_0}\exp[-\beta{\cal H}]}
{{\rm Z}(J)}
\sim \exp[-r\epsilon(J)^{\nu(J)}]\label{c},
\eeq
and also we study the symmetry of the susceptibility 
\beq
\chi(J)\equiv\sum_{i_0j_0}\omega_{i_0j_0}(J)
\sim \epsilon(J)^{-\gamma(J)},\label{gamma}
\eeq
where $\{g_{i_0j_0}\}$ are the sign 
corresponding to the emerging order 
and $r$ is the distance between the site $i_0$ and $j_0$. 
We can assume without loss of generality that 
the site $i_0$ lies on ${\cal H}_1$. 
Corresponding to the change of the interaction $J$ to $-J$, 
the quantity $g_{i_0j_0}$ changes its sign 
when the site $j_0$ belongs to ${\cal H}_2$ 
and then $g_{i_0j_0}s_{i_0}s_{j_0}$ is even 
for the spin invergion of ${\cal H}_2$. 
As a result, we again arrive at 
\beq
\omega_{i_0j_0}(J)=\omega_{i_0j_0}(-J)\ \ {\rm and}\ \ 
\chi(J)=\chi(-J),
\eeq
and hence
\beq
\nu(J)=\nu(-J)\ \ {\rm and }\ \ 
\gamma(J)=\gamma(-J).\label{gamsym}
\eeq

\subsection{Vanishing cases}
The above argument cannot exclude the case that 
the critical temperature 
and exponents do not differentiable at $J=0$. 
The critical coefficient, indeed, 
behaves as a cusp and do not differentiable 
at $J=0$ 
when the system is not translationally invariant. 
Here we show using the scaling hypothesis  
that the first-order derivatives 
of the critical temperature and exponents 
exist and they are vanishing. 
Let us consider the free energy $f$ 
without the external field as 
\beq
f\simeq C\epsilon^{2-\alpha}\log\epsilon, 
\label{freeng}
\eeq
where $C$ is some constant. 
Differentiating (\ref{freeng}) 
with respect to the interaction $J$ and 
taking a limit $J\rightarrow 0$, we obtain 
\beqa
0&=&(\frac{\partial C}{\partial J})_0 
\epsilon^2\log\epsilon 
+C\epsilon^2\frac{1}{\epsilon}
(\frac{\partial\epsilon}{\partial J})_0 
\nonumber\\
&+&C\epsilon^2\log\epsilon
[(2-\alpha)\frac{1}{\epsilon}
(\frac{\partial\epsilon}{\partial J})_0
-(\frac{\partial\alpha}{\partial J})_0\log\epsilon] 
\eeqa
and as a result 
\beq
(\frac{\partial C}{\partial J})_0=0, \ 
(\frac{\partial\alpha}{\partial J})_0=0 \ {\rm and}\ 
(\frac{\partial\epsilon}{\partial J})_0=0. 
\label{derzeroae}
\eeq

Next let us consider the free energy $f$ 
with the external field $h$. 
We assume the scaling form of $f$ as 
\beq
f_J(\lambda^p\epsilon,\lambda^qh)=\lambda f_J(\epsilon,h), 
\label{freengsc}
\eeq
where $\lambda$ is some parameter and 
$p$ and $q$ are numbers independent on $\epsilon$ and $h$. 
They are related to the exponents $\alpha$ and $\gamma$ as 
\beq
\alpha=2-\frac{1}{p} \ {\rm and}\  \gamma=\frac{2q-1}{p}. 
\label{exppar}
\eeq
From (\ref{derzeroae}) and (\ref{exppar}), we obtain 
$(\partial p/\partial J)_0=0$. 
The partial derivative of $f$ shows the same scaling form as 
\beq
\frac{\partial f_J(\lambda^p\epsilon,\lambda^qh)}{\partial J}
=\lambda\frac{\partial f_J(\epsilon,h)}{\partial J}.
\eeq
Differentiating (\ref{freengsc}) with respect to $J$, 
a straightforward calculation yields 
\beq
M(0,1)\frac{\log h}{q}
(\frac{\partial q}{\partial J})_{J=0}=0,
\label{condfin}
\eeq
for  $\epsilon\rightarrow 0$, $J\rightarrow 0$ 
and $\lambda^qh=1$, 
where $M(\epsilon, h)=\partial f_J(\epsilon, h)/\partial
h$ is  the magnetization. 
Note that $(\partial q/\partial J)$ is independent on $h$. 
From (\ref{condfin}), we obtain 
$(\partial q/\partial J)_{J=0}=0$, 
and as a result
\beq
(\frac{\partial \gamma}{\partial J})_{J=0}=0. 
\label{derzerobdg}
\eeq
 
\section{The nonvanishing derivatives}
\setcounter{equation}{0}
In this section, 
we show the non-universal behaviour of the relevant system 
when it satisfies the conditions 1) and 2). 
We use the formulas (\ref{gencor}) and (\ref{grncan}) 
derived from 1) and 2), 
respectively, 
and (\ref{derzeroae}) and (\ref{derzerobdg}) 
result from the symmetry of the model. 
The properties 
(\ref{grncan}), (\ref{derzeroae}) and (\ref{derzerobdg}) 
are valid when $n^{(k)}$ or $n^{(l)}$ 
(or both $n^{(k)}$ and $n^{(l)}$) are odd. 
Let us consider the weighted susceptibility 
\beq
\chi=\beta\mu_B^2
\sum_{i_0j_0}g_{i_0j_0}\langle s_{i_0}s_{j_0}\rangle, 
\label{weisus}
\eeq
where $g_{i_0j_0}$ is the sign 
corresponding to the emerging order as 
$g_{i_0j_0}={\rm sgn}(g_{i_0}g_{j_0})$. 
All we have to do here is to differentiate (\ref{weisus}) 
in terms of $J_{kl}$ 
and to show that the second dominant term 
shows the logarithmic singularity 
$\chi_0\log\epsilon$ with $\epsilon=(T-T_{\rm c})/T_{\rm c}$. 

Let us introduce the following notations. 
The interaction ${\cal H}_{kl}$ is expressed as 
\beq
{\cal H}_{kl}=-J\sum_{i}{\cal O}_{i},\ \ {\rm and}\ \ 
{\cal O}_{i}={\cal O}_{i}^{(k)}{\cal O}_{i}^{(l)}, 
\eeq
where $J_{kl}$ is written as $J$ for simplicity. 
The vector $r_i$ denotes the coordinate of the spin $s_i$ and 
$R_i=\{r_{i_1}, \cdots, r_{i_n}\}$ 
denotes the coordinate  of the spin product 
${\cal O}_i=s_{i_1}\cdots s_{i_n}$. 
The expectation value of ${\cal O}_i$ 
is expressed as a function of $R_i$ as 
\beq
\langle {\cal O}_i\rangle=
c_0(R_i)+c_1(R_i)\epsilon\log\epsilon. 
\eeq
We also write $R_i\subset {\cal R}_m$ (or $i\in{\cal R}_m$) 
when $r_s \in {\cal R}_m$ for all $r_s \in R_i$. 

Here we derive finite derivatives of the exponent $\gamma$. 
The first-order derivative of (\ref{weisus}) is 
\beq
(\frac{\partial\chi}{\partial J})_{J=0}=
\beta^2\mu_{\rm B}^2\sum_{i_0j_0}g_{i_0j_0}
\sum_{i}[\langle s_{i_0}s_{j_0}
        {\cal O}_{i}^{(k)}{\cal O}_{i}^{(l)}\rangle_0-
\langle s_{i_0}s_{j_0}\rangle_0
\langle {\cal O}_{i}^{(k)}{\cal O}_{i}^{(l)}\rangle_0],
\label{chider1}
\eeq
where $\langle\ \rangle_0$ 
denotes the expectation value taken for $J=0$. 
All the terms cancel except the cases: 

\bdes
\item[I-1)]
$s_{i_0}$, $s_{j_0}\in{\cal H}_k$ 
(or $s_{i_0}$, $s_{j_0}\in{\cal H}_l$), 
and both $n^{(k)}$ and $n^{(l)}$ are even, and 
\item[I-2)]
$s_{i_0}\in{\cal H}_k$, $s_{j_0}\in{\cal H}_l$, 
and both $n^{(k)}$ and $n^{(l)}$ are odd. 
\edes
The derivatives $(\partial\gamma/\partial J)_0$, 
$(\partial T_{\rm c}/\partial J)_0$ 
of the latter case I-2) vanish 
as shown in section 6. 
As a result, 
we have only to treat the case I-1) 
as the first-order derivative. 
This case is a simple generalization 
of the argument for the eight-vertex 
model in section 4. 
The derivative (\ref{chider1}) is written as 
\beq
(\frac{\partial\chi}{\partial J})_{J=0}=
\beta^2\mu_{\rm B}^2\sum_{i_0j_0}g_{i_0j_0}
\sum_{i}[\langle s_{i_0}s_{j_0}{\cal O}_{i}^{(k)}\rangle_0-
\langle s_{i_0}s_{j_0}\rangle
\langle{\cal O}_{i}^{(k)}\rangle_0]
\langle{\cal O}_{i}^{(l)}\rangle_0.\label{chider1d}
\eeq
This is a generalization of (\ref{dchi8V}). 
From (\ref{dchidif}) and (\ref{gencor}), 
we obtain the following temperature dependence 
\beq
(\frac{\partial\chi}{\partial J})_{J=0}\simeq
\chi_0\sum_i{\cal G}_i
[a_0(R_i)\frac{1}{\epsilon}+a_1(R_i)\log\epsilon]
[c_0(R_i)+c_1(R_i)\epsilon\log\epsilon],
\label{derfrst}
\eeq
where $a_0(R_i)$, $a_1(R_i)$, $c_0(R_i)$ and $c_1(R_i)$ 
are some constants. 
The term ${\cal G}_i={\cal G}^{(k)}_i{\cal G}^{(l)}_i$ 
is factorized 
so that the inside of the brackets $[\cdots][\cdots]$ 
in (\ref{derfrst}) 
is positive. 
We generally cannot omit the term $a_1(R_i)$. 
The sum $\sum_i {\cal G}_i^{(k)}a_0(R_i)$ 
and $\sum_i {\cal G}_i^{(k)}a_1(R_i)$ 
appear as the coefficient of 
$(\partial{\tilde\chi}_k/\partial{\tilde J}_{k})_{{\tilde J}_k=0}$ 
where ${\tilde\chi}_k$ is the susceptibility of the model 
described by the Hamiltonian 
${\tilde{\cal H}}_k={\cal H}_k-{\tilde J}_k\sum_i{\cal O}^{(k)}_i$ 
and they are finite in the thermodynamic limit. 
The coefficient of $\log\epsilon$, namely 
\beq
\sum_i {\cal G}_i(a_0(R_i)c_1(R_i)+a_1(R_i)c_0(R_i)), 
\label{psdsus}
\eeq
is finite because $\sum_i {\cal G}_i^{(k)}a_0(R_i)$, 
$\sum_i {\cal G}_i^{(k)}a_1(R_i)$, 
${\cal G}_i^{(l)}c_0(R_i)$ 
and ${\cal G}_i^{(l)}c_1(R_i)$ are all finite. 
This is the first order derivative of $\gamma$. 

For all the cases except when both 
${\cal O}_{i}^{(k)}$ and ${\cal O}_{i}^{(l)}$ 
have even symmetry 
(i.e. the case I-1: both $n^{(k)}$ and $n^{(l)}$ are even), 
we obtained 
from (\ref{derzeroae}) and (\ref{derzerobdg})  
\beq
(\frac{\partial T_{\rm c}}{\partial J})_0=0\ {\rm and}\ 
(\frac{\partial \gamma}{\partial J})_0=0. 
\label{derzero} 
\eeq 
In these cases, 
we have to show that the second-order derivatives 
are non-vanishing  and finite. 
From (\ref{dchieps}) and (\ref{derzero}), 
the second-order derivative of 
the susceptibility $\chi$ is 
\beq
(\frac{\partial^2\chi}{\partial J^2})_{J=0}\simeq\chi_0[
-\gamma(0)\frac{1}{\epsilon(0)}
(\frac{\partial^2\epsilon}{\partial J^2})_0
-(\frac{\partial^2\gamma}{\partial J^2})_0\log\epsilon(0)]. 
\eeq
Hence the existence of the logarithmic singularity is 
the sign of continuously varying critical exponents. 
All we have to do is 
to find a term proportional to $\chi_0\log\epsilon$ 
in the second-order derivative of $\chi$ 
and to show the coefficient of $\chi_0\log\epsilon$ is finite. 

Differentiating 
$\chi=\beta\mu_{\rm B}^2\sum\langle s_{i_0}s_{j_0}\rangle$ 
with respect to $J$ twice, we obtain 
\beq
(\frac{\partial^2\chi}{\partial J^2})_{J=0}=
\beta^3\mu_{\rm B}^2\sum_{i_0j_0}g_{i_0j_0}
\sum_{i}\sum_{j}
[\langle s_{i_0}s_{j_0}
{\cal O}_{i}^{(k)}{\cal O}_{j}^{(k)}\rangle_0-
\langle s_{i_0}s_{j_0}\rangle_0
\langle{\cal O}_{i}^{(k)}{\cal O}_{j}^{(k)}\rangle_0]
\langle{\cal O}_{i}^{(l)}{\cal O}_{j}^{(l)}\rangle_0,
\label{chider2}
\eeq
where we have used that $n^{(k)}$ or $n^{(l)}$ 
(or both $n^{(k)}$ and $n^{(l)}$) are odd and 
the expectation value 
$\langle s_{i_1}\cdots s_{i_n}\rangle_0$ 
equals zero 
when $n$ is odd. 
The following case remains nonvanishing:
\bdes
\item[II-1)]
$s_{i_0}$, $s_{j_0}\in{\cal H}_k$ 
(or $s_{i_0}$, $s_{j_0}\in{\cal H}_l$), 
$n^{(k)}$ is even and $n^{(l)}$ is odd, and 
\item[II-2)]
$s_{i_0}$, $s_{j_0}\in{\cal H}_k$ (or $s_{i_0}$, 
$s_{j_0}\in{\cal H}_l$), 
and both $n^{(k)}$ and $n^{(l)}$ are odd. 
\edes
Both cases can be treated simultaneously. 
From (\ref{dchidif}) and (\ref{gencor}), 
the derivative (\ref{chider2}) 
shows the following temperature dependence 
\beqa
&&\chi_0\sum_i\sum_j{\cal G}_i{\cal G}_j
[a_0(R_i,R_j)\frac{1}{\epsilon}+a_1(R_i,R_j)\log\epsilon]
[c_0(R_i,R_j)+c_1(R_i,R_j)\epsilon\log\epsilon],\nonumber\\
&\equiv& \chi_0\sum_i\sum_j{\cal G}_i{\cal G}_j\Omega(R_i, R_j), 
\eeqa
where $\Omega(R_i, R_j)$ is positive and  
$a_0(R_i,R_j)$, $a_1(R_i,R_j)$, 
$c_0(R_i,R_j)$ and $c_1(R_i,R_j)$ 
are some constants. 
This summation can be regrouped 
by $\{{\cal R}_m\}$ and using the condition 
(\ref{grncan}) 
(i.e. the condition 2: $\sum_{i\in{\cal R}_m}{\cal G}_i=0$), 
we obtain 
\beqa
&&\chi_0\sum_{mm'}\sum_{i\in{\cal R}_m}\sum_{j\in{\cal R}_{m'}}
{\cal G}_i{\cal G}_j
\Omega(R_m+\Delta R_{mi}, R_{m'}+\Delta R_{m'j})\nonumber\\
&\simeq&\chi_0\sum_{mm'}\sum_{i\in{\cal R}_mj\in{\cal R}_{m'}}
{\cal G}_i{\cal G}_j  (\Delta R_{mi}\cdot\nabla)(\Delta
R_{m'j}\cdot\nabla')\Omega(R_m, R_{m'}), \label{coefexp}
\eeqa
where $R_m\subset{\cal R}_m$ and $R_{m'}\subset{\cal R}_{m'}$ 
are fixed for each 
${\cal R}_m$ and ${\cal R}_{m'}$, respectively, and  
we have used the following notation 
\beqa
\Delta R_{mi}\cdot\nabla&\equiv&
\Delta r_{m_1i_1}\cdot\nabla_{i_1}
+\cdots+\Delta r_{m_ni_n}\cdot\nabla_{i_n} \nonumber\\
\Delta R_{m'j}\cdot\nabla'&\equiv&
\Delta r_{m'_1j_1}\cdot\nabla'_{j_1}
+\cdots+\Delta r_{m'_nj_n}\cdot\nabla'_{j_n}
\eeqa
where $\nabla_i$ and $\nabla'_i$ are the gradient operating 
to the coordinate of $s_i$ 
and $\Delta_{mi}\cdot\nabla$ 
and $\Delta_{m'j}\cdot\nabla'$ operate on 
the first and the second arguments 
of $\Omega(R_m, R_{m'})$. 

The first summation in (\ref{coefexp}) 
is classified by the distance 
between ${\cal R}_m$ and ${\cal R}_{m'}$ as 
\beq
\sum_{mm'}=\sum_r\sum_{mm'|m-m'|=r},
\eeq
where $|m-m'|=r$ denotes that 
$r\leq{\rm min}\{|r_i-r_{i'}|\ 
|\ r_i\in{\cal R}_m, r_{i'}\in{\cal R}_{m'}\}<r+\Delta r$ 
and $\Delta r$ is a constatnt 
comparable to the mean size of $\{{\cal R}_m\}$. 

As $\Omega$ is a smooth and decreasing function of $r$ 
and all $\{{\cal R}_m\}$ 
denote finite regions containing a finite number of spins, 
each term in (\ref{coefexp}) is bounded by 
\beq
{\cal G}_i{\cal G}_j
(\Delta R_{mi}\cdot\nabla)(\Delta R_{m'j}\cdot\nabla')\Omega\leq
{\rm const.}(\Delta R\cdot\nabla)(\Delta R\cdot\nabla')\Omega,
\eeq
where
\beq
\Delta R\cdot\nabla \equiv
\Delta r\cdot\nabla_{i_1}+\cdots+\Delta r\cdot\nabla_{i_n},
\eeq
and a similar equation is defined for $\Delta R\cdot\nabla'$. 

Then the coefficient of $\chi_0\log\epsilon$ in (\ref{coefexp}) 
is bounded as 
\beqa
&&\sum_r 
(\Delta R_{mi}\cdot\nabla)(\Delta R_{m'j}\cdot\nabla')
F(r,\{R_i\})\nonumber\\
&<&\sum_{r<r_0}
(\Delta R_{mi}\cdot\nabla)(\Delta R_{m'j}\cdot\nabla')
F(r,\{R_i\})
+{\rm const.}\int_{r_0}^{\infty}{\rm d}^2r 
(\Delta R\cdot\nabla)(\Delta R\cdot\nabla')
F(r,\{R_m\}),\nonumber\\
\label{coefln} 
\eeqa
where
\beq
F(r,\{R_i\})
=\sum_{mm'|m-m'|=r}\sum_{i\in{\cal R}_mj\in{\cal R}_{m'}}
{\cal G}_i{\cal G}_j
[a_0(R_i,R_j)c_1(R_i,R_j)+a_1(R_i,R_j)c_0(R_i,R_j)]
\eeq
is a finite function of $r$ (see (\ref{psdsus})). 

Our purpose is to show (\ref{coefln}) is finite. 
The first sum is of course finite 
and the second term is bounded by 
\beqa
A\int_{r_0}^{\infty}{\rm d}^2r \nabla^2 F(r,\{R_i\})
=A\int_{r=r_0}{\rm d}S \frac{\partial F}{\partial r}
\simeq A\cdot 2\pi r_0 \frac{\partial F}{\partial r}|_{r=r_0}
<\infty, 
\label{finite}
\eeqa
where $A$ is some constant.

\section{Generalization}
\setcounter{equation}{0}
The condition 2) can be generalized 
in the following two points. 

At first, 
it is straightforwald to generalize the form of the interaction 
${\cal H}_{kl}$ as
\beq
{\cal H}_{kl}=\sum_p{\cal H}_{kl}^{(p)}\ \ {\rm and}\ \ 
{\cal H}_{kl}^{(p)}=
-J_{kl}^{(p)}\sum{\cal O}_{pi}^{(k)}{\cal O}_{pi}^{(l)},
\eeq
where each ${\cal H}_{kl}^{(p)}$ satisfies the condition 2). 

Next, 
the values of $n^{(k)}$ and $n^{(l)}$ have been fixed. 
This condition is necessary 
for showing the cancellations of correlations in 
the  zeroth- and the first-order terms 
in (\ref{coefexp}) using the condition 
$\sum_{i\in{\cal R}_m}{\cal G}_i=0$. 
However, 
our argument is valid for more complicated interactions 
i.e. the case that $n^{(k)}$ and $n^{(l)}$ 
depend on the region ${\cal R}_m$. 
Each contribution from each ${\cal R}_m$ to the derivatives 
is classified into the cases shown in sections 6 and 7. 
As for the first-order derivatives, there is no difference. 
For the second-order derivative, all the contributions 
to $(\partial^2\gamma/\partial J^2)_0$ 
coming from ${\cal R}_m$ and ${\cal R}_{m'}$ vanish 
except when both 
$n^{(k)}_m+n^{(k)}_{m'}$ and $n^{(l)}_m+n^{(l)}_{m'}$ 
are even. 
(Otherwise the argument in section 6 is valid for each term and 
the contributions to the derivative is vanishing.) 
The nonvanishing case can be treated 
in the same way as in section 7.

\section{Conclusion}
\setcounter{equation}{0}
We have derived that 
some Ising systems satisfying the conditions 1) and 2) 
in section 2 show non-universal critical behaviour. 
The perturbational expansion in terms of the interaction $J$ 
is performed. 
This method results in the exact first-order derivatives 
(\ref{dgam}) 
in the case of the eight-vertex model. 
We have used the conditions (\ref{gencor}) and (\ref{grncan}), 
which is derived from the conditions 1) and 2), respectively, 
and (\ref{derzeroae}) and (\ref{derzerobdg})
which result from the symmetry of the model. 
The existence of finite and non-zero derivatives 
$(\partial\gamma/\partial J)_0$ 
or $(\partial^2\gamma/\partial J^2)_0$ 
is an evidence of continuous variation 
of critical exponents. 
These derivatives are derived 
in (\ref{chider1})-(\ref{psdsus}) 
and (\ref{chider2})-(\ref{finite}) 
for the first- and the second-order derivatives, respectively. 
Finally, 
the straightforward generalization of the condition 2) 
is commented in section 8. 
This condition is valid for generalized spin S Ising models 
or can easily be generalized for other classical systems.

$\;$

\noindent
{\bf\Large Acknowledgements}

$\;$

It is our great pleasure 
to express our gratitude to Dr.T.Chikyu 
for his useful comments. 


\end{document}